# Synthesis and characteristion of $Li_xBC$ – hole doping does not induce superconductivity

**A.M. Fogg, J.B. Claridge, G.R. Darling and M.J. Rosseinsky***
*Department of Chemistry, University of Liverpool, Liverpool L69 7ZD, UK*

**The synthesis of $Li_xBC$ (x > 0.5) by high temperature Li deintercalation from LiBC is demonstrated by refinement of X-ray and neutron powder diffraction data – contrary to theoretical expectation no superconductivity above 2K is observed in these materials.**

The layered diboride $MgB_2$[1] is superconducting at 39K. It adopts an AAA stacking of graphite-like $B_2$ sheets with Mg in the interlayer space. The superconductivity in $MgB_2$ is attributed to hole carriers arising from an unusual "self-doping" mechanism in which the stabilising effect of the interlayer Mg on the B $2p_\pi$ states in the layer pushes them down in energy to overlap the in-plane σ bonding states and generate holes in these states at the Fermi energy.[2]

LiBC is the closest known analogue of $MgB_2$: it is isoelectronic and adopts a related structure (Figure 2(i) inset) in which the hexagonal graphite-like sheets persist but now contain strict alternation of B and C on the vertices.[3] This alternation is maintained in the layer stacking motif which gives an AB stacking, with B superposed directly on C. Band structure calculations[4,5] indicate the opening of a gap at the Fermi level in a density of states otherwise very similar to $MgB_2$ due to the observed strict B-C alternation between the layers.

Due to the role of hole-doping in rendering $MgB_2$ metallic, attention has recently focussed on the possibility of achieving such doping in LiBC. The structural similarity with the graphite intercalation compounds suggests that it may be possible to prepare $Li_xBC$ phases with partial occupancy of the interlayer sites. This would correspond to oxidation of the $(BC)^-$ layer and the introduction of hole carriers, producing an analogy with the electronic structure of $MgB_2$. These hole-doped $Li_xBC$ phases are now important synthetic targets as recent detailed calculations suggest that these phases will have even higher $T_c$'s than found in $MgB_2$ itself.[4] Attempts to investigate this in detail have been confounded by the difficulty in locating and quantifying Li in poorly crystalline multiphase $Li_xBC$ assemblages.[6]

Here we describe the synthesis of a family of heavily hole-doped derivatives of LiBC, with Li deficiencies of over 50%. These materials are crystalline and analysed quantitatively by neutron and X-ray powder diffraction to allow evaluation of the Li concentration. Despite the large formal hole concentration in the π bands, no superconductivity is observed at above 2K.

The original report of LiBC synthesis involves reaction at 1500 °C for one hour.[3] Extended treatment at this temperature, for up to 24 hours, results in loss of lithium but retention of the LiBC structure. Laboratory X-ray powder diffraction reveals that the unit cell volume decreases with increasing time at temperature. Rietveld refinement of powder neutron and synchrotron X-ray diffraction (Figure 1(i)) data indicate that the decrease in unit cell volume occurs together with refineable loss of lithium from the interlayer sites while the layer B/C sites retain complete occupancy. The retention of the 101 reflection demonstrates that the AB layer stacking with strict B/C alternation is retained. High resolution diffraction data reveal phase separation in the strongly Li-deficient phases. The refined data shown in Figure 1(i) are from a sample of overall composition $Li_{0.78(1)}BC$ prepared by 24 hour treatment at 1500 °C (using $^{11}B$ - see ESI) which consists of $Li_{0.56(1)}BC$ ($a$ = 2.62631(7)Å, $c$ = 7.4390(4)Å) and $Li_{0.90(1)}BC$ ($a$ = 2.72610(3)Å, $c$ = 7.1112(1)Å) compared with $a$ = 2.75009(1)Å and $c$ = 7.05376(6)Å for LiBC, all refined in space group $P6_3/mmc$. A single phase sample prepared at the same temperature and reaction time (using natural B) refines to a Li concentration of $Li_{0.518(7)}BC$. The unit cell volume variation with x is shown in Figure 2(i) and that of the B-C and Li-B/C distances in Figure 2(ii).

The volume-composition data show a group of phases with Li contents greater than 0.8 followed by a sharp contraction in the unit cell volume accompanied by a rapid decrease in the refined Li content. This is suggestive of solid solution for Li > 0.8 followed by a two phase region and either a second solid solution or series of line phases at lower x content below 0.6. The cell volume decreases with Li loss due to the reduced guest volume. The B-C distances (Figure 2(ii)) in the most strongly oxidised $Li_{0.5+x}BC$ systems are notably short – 1.51496Å

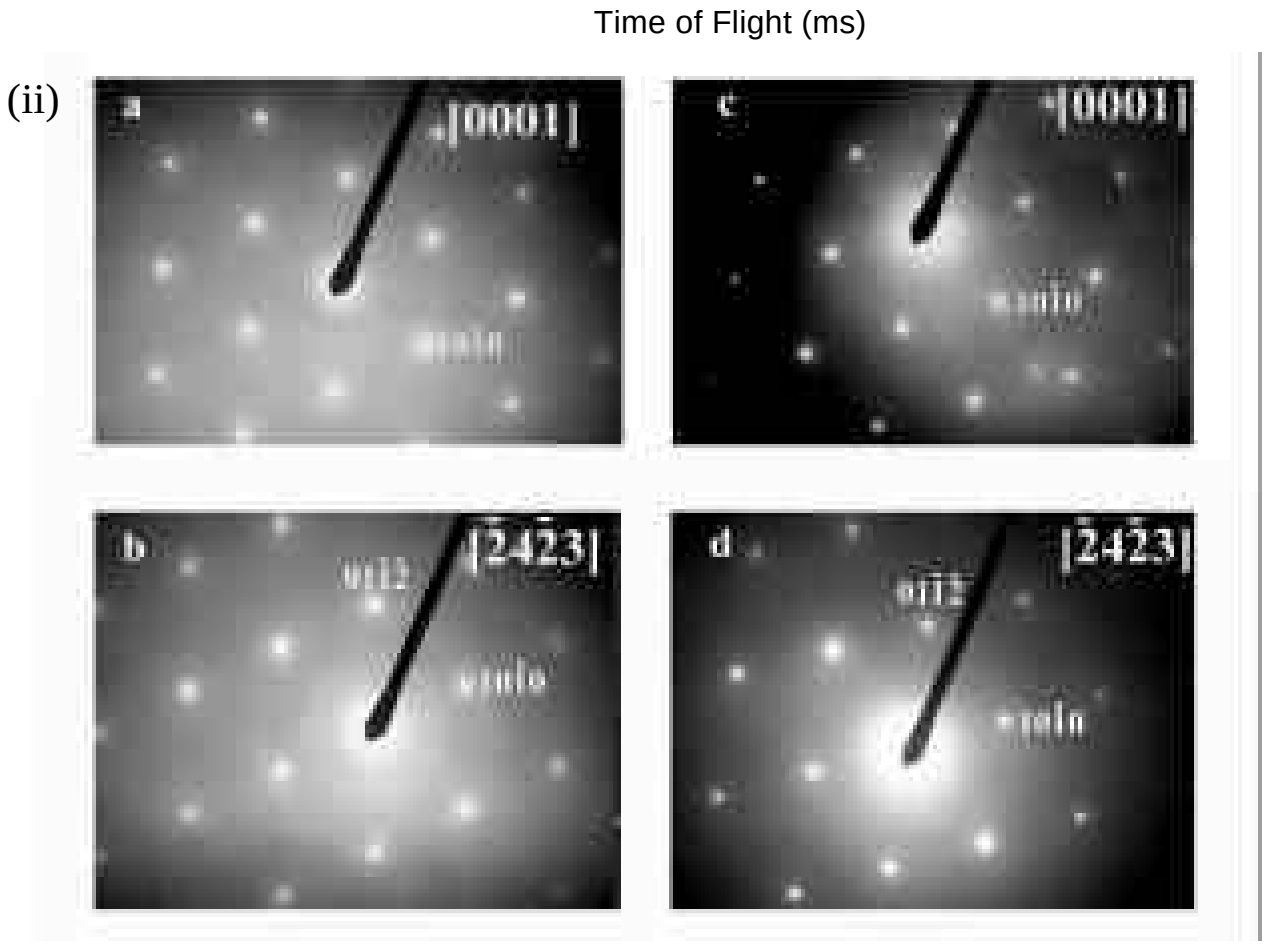

**Fig. 1** (i) Rietveld refinement of $Li_{0.78(1)}BC$ prepared at 1500 °C for 24 hours. The two phases present are $Li_{0.56(1)}BC$ (lower tick marks, 35.2(3)wt%) and $Li_{0.90(1)}BC$. Data shown are for the backscattering bank (2θ = 154°; d = 0.4 – 2.0Å) of GEM; $R_{wp}$ = 4.39%, $R_e$ = 3.14%. (ii) SAED patterns for the $Li_{0.56(1)}BC$ ((a) and (b)) and $Li_{0.90(1)}BC$ ((c) and (d)) phase components of the sample refined in Figure 1(i) above. This shows the absence of super cell reflections in both phases. The ratio of the lattice constants of the two phases agrees with that refined from powder neutron diffraction, vindicating the two-phase analysis of the sample.

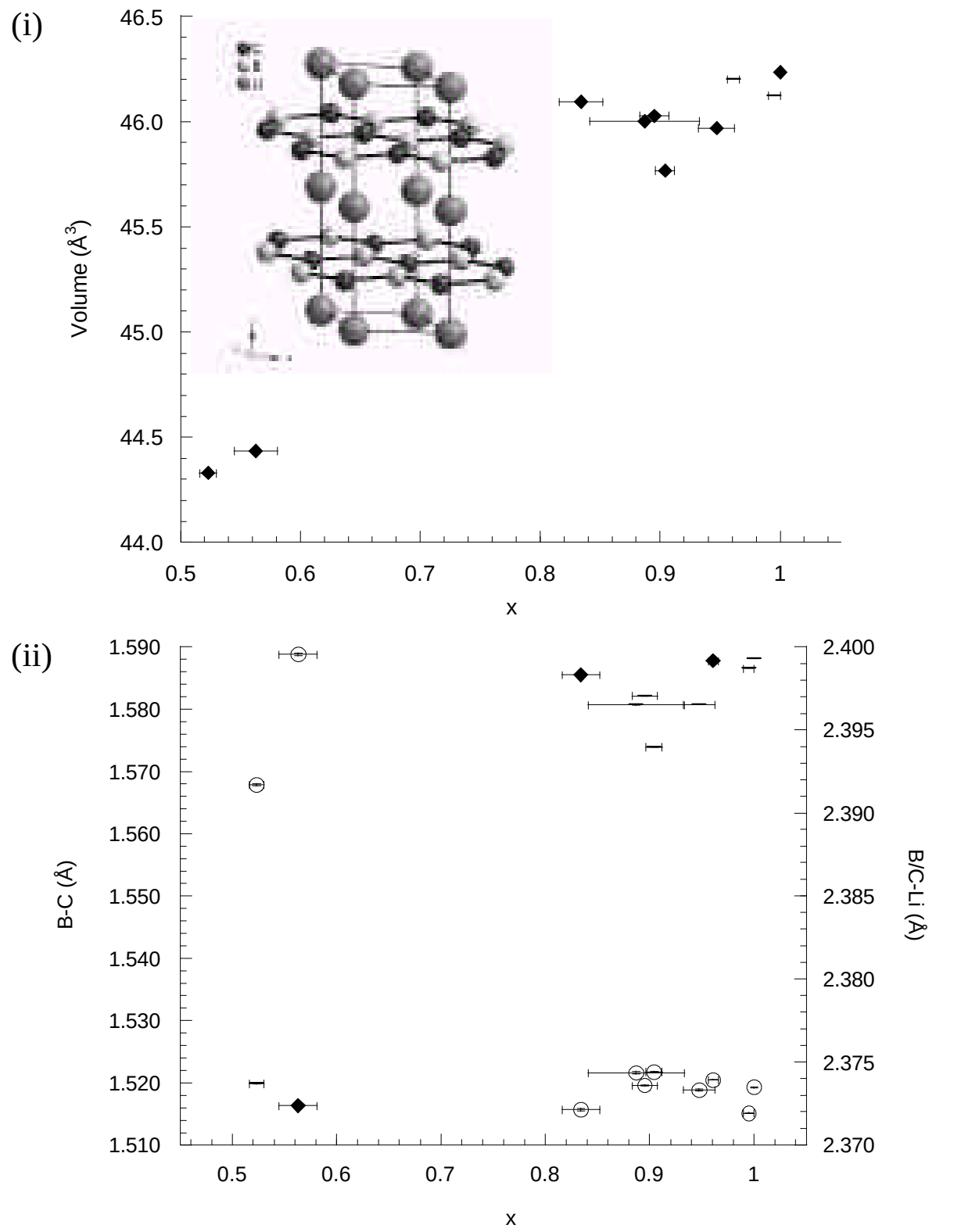

**Fig. 2** (i) Variation of unit cell volume with refined x values Inset: The layered structure of LiBC. (ii) Variation of intralayer B-C distance (filled diamonds) and interlayer Li-B/C distance (hollow circles) with Li content.

corresponds to a 0.07Å shortening of this distance on oxidation. This agrees well with the results of DFT geometry optimisation of different supercell structures with stoichiometry $Li_{0.5}BC$. The B-C distance shortens by 0.02-0.04Å compared to the optimized 1.566 A for unoxidised LiBC. This shortening of the bond is likely due to the reduced Coulomb repulsion of the B and C atoms, since the charge on each (determined by Mulliken population analysis) is 0.1-0.2 electrons less than for LiBC (for which the charge on B is -0.21 e and on C -0.72 e). In contrast to the contraction in the volume and in *a*, the interlayer distance *c* and the Li-B/C distances (Figure 2(ii)) increase markedly on oxidation, again in agreement with DFT optimisation. This is presumably due to reduced electrostatic interaction as the layer becomes less negatively charged. The 4.5% decrease in the B-C distance overcomes the 1.1% increase in the Li-B/C distance on oxidation to drive the contraction of the unit cell.

The absence of superstructure formation due to Li vacancy ordering within or between the planes is surprising but is confirmed by electron diffraction (Figure 1(ii)). Removal of Li would be expected to produce pronounced relaxation of the structure in the immediate environment of the vacancy as the BC layer responds to the different charge in this region. Incoherent local displacements will contribute to the Debye-Waller factors of B and C: these become more anisotropic due to out-of-plane displacements as the level of oxidation increases. The rms B/C out-of-plane displacements at 300K and 5K for x = 0.56 are 0.15Å and 0.12Å respectively. The small decrease in this out -of-plane displacement on cooling to 5K indicates that a large component is static reflecting local relaxation of the B and C atoms in the sheet in the vicinity of an Li vacancy.

Although the suggested level of oxidation of the $BC^-$ layers required for superconductivity is achieved here, d.c. magnetisation measurements demonstrate the absence of superconductivity above 2K in all the $Li_xBC$ materials reported here. Superconductivity was originally postulated in hole-doped LiBC on the basis of a comparison of the electronic structure to that of $MgB_2$. While in LiBC, the B-C sigma states (similar to those responsible for superconductivity in $MgB_2$) are fully occupied lying below $E_f$, when less charge is transferred into the B-C layers, these bands shift rigidly up (by about 1.5 eV for $Li_{0.5}BC$) to lie across the Fermi energy, as shown in Fig. 3. That superconductivity does not result may be due to disorder in the layers. Geometry optimisation of $Li_{0.5}BC$ yields structures in which the B-C bond lengths can vary by as much 0.04**Å**. Layers can also be very noticeably buckled by as much as 0.34Å, consistent with the enhanced out-of-plane displacement parameters seen on oxidation. In contrast we find no layer buckling in the DFT geometry optimisation of LiBC, even when starting from highly distorted supercell structures, in agreement with the planar structure derived from diffraction. Such strong local variation of the B-C distances and the buckling of the layers (which must be local and uncorrelated over even electron diffraction length scales) will lead to a mixing of π and σ states in $Li_xBC$ that may be strong enough to destroy the superconductivity. Although the density of states in LiBC and its oxidised $Li_xBC$ derivatives appears to indicate rigid band doping, the local electronic structure differs strongly. The absence of superconductivity may have its origins in strong relaxation of the structure in the vicinity of the dopants, but further detailed investigation of the normal state both computationally and experimentally is required for detailed comment. The synthesis of Li-deficient derivatives of LiBC is chemically significant as it opens a new avenue for controlling the electronic properties of $MgB_2$-related materials with electronic structures derived from graphite.

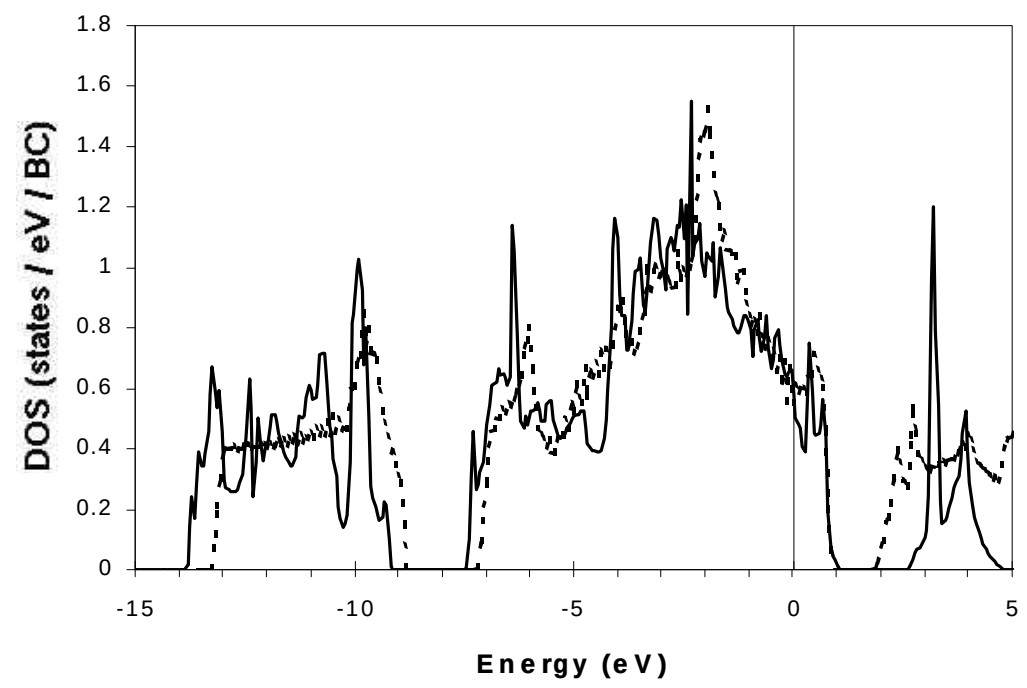

**Fig. 3** The density of states (dos) in geometry-optimised $Li_{0.5}BC$ (solid line) and in LiBC (dashed line). The dos in LiBC has been shifted upwards in energy by 1.5 eV.. The Fermi energy of $Li_{0.5}BC$ is zero on the energy scale and marked with a vertical line.

## Notes and references

Full details are given in ESI. LiBC was prepared as in 3 in a sealed Ta tube and then held at the synthesis temperature of 1773K for between one and 24 hours, with increased Li loss resulting from longer hold times. Samples for neutron diffraction were prepared in a similar manner from $^{11}B$ ( Eagle-Picher Technologies). d.c. magnetisation measurements were performed in 20G and 40G fields in a Quantum Design MPMS XL-7 SQUID magnetometer. Neutron powder diffraction data were collected on the GEM instrument at the ISIS spallation neutron source, Rutherford Appleton Laboratory at 300K and 5K. Selected area electron diffraction patterns were obtained with of a double-tilting goniometer stage (±30°) in a JEOL 2000 FX transmission electron microscope. DFT calculations have been performed with a plane-wave basis set, with an energy cut-off of 340 eV, using ultra-soft pseudopotentials for the atoms. For the $Li_xBC$ structures, the geometry was optimized in P1 symmetry using a Brillouin zone sampling of 24-27 special k-points. Exchange and correlation has been included using the PBE version of GGA. We thank EPSRC for funding under GR/R53999.